\documentclass[reprint,twocolumn,aps,prl,floats,superscriptaddress]{revtex4-1}
\usepackage[english]{babel}
\usepackage[ansinew]{inputenc}
\usepackage{amsmath}
\usepackage{amssymb}
\usepackage{color}
\usepackage{graphicx}

\begin{document}
\title{Interplay between Orbital Magnetic Moment and Crystal Field Symmetry: \\ Fe atoms on MgO}

\author{S.~ Baumann}
\affiliation{IBM Almaden Research Center, 650 Harry Road, San Jose, CA 95120, USA}
\affiliation{Department of Physics, University of Basel, Klingelbergstrasse 82, CH-4056 Basel, Switzerland}
\author{F.~Donati}
\affiliation{Institute of Condensed Matter Physics, {\'E}cole Polytechnique F{\'e}d{\'e}rale de Lausanne (EPFL), Station 3, CH-1015-Lausanne, Switzerland}
\author{S.~Stepanow}
\affiliation{Department of Materials, ETH Z{\"u}rich H{\"o}nggerbergring 64, CH-8093 Z{\"u}rich, Switzerland}
\author{S.~Rusponi}
\affiliation{Institute of Condensed Matter Physics, {\'E}cole Polytechnique F{\'e}d{\'e}rale de Lausanne (EPFL), Station 3, CH-1015-Lausanne, Switzerland}
\author{W. Paul}
\affiliation{IBM Almaden Research Center, 650 Harry Road, San Jose, CA 95120, USA}
\author{S.~ Gangopadhyay}
\affiliation{IBM Almaden Research Center, 650 Harry Road, San Jose, CA 95120, USA}
\author{I.~ G.~ Rau}
\affiliation{IBM Almaden Research Center, 650 Harry Road, San Jose, CA 95120, USA}
\author{G.~E.~ Pacchioni}
\affiliation{Institute of Condensed Matter Physics, {\'E}cole Polytechnique F{\'e}d{\'e}rale de Lausanne (EPFL), Station 3, CH-1015-Lausanne, Switzerland}
\author{L.~Gragnaniello}
\affiliation{Institute of Condensed Matter Physics, {\'E}cole Polytechnique F{\'e}d{\'e}rale de Lausanne (EPFL), Station 3, CH-1015-Lausanne, Switzerland}
\author{M.~Pivetta}
\affiliation{Institute of Condensed Matter Physics, {\'E}cole Polytechnique F{\'e}d{\'e}rale de Lausanne (EPFL), Station 3, CH-1015-Lausanne, Switzerland}
\author{J.~Dreiser}
\affiliation{Institute of Condensed Matter Physics, {\'E}cole Polytechnique F{\'e}d{\'e}rale de Lausanne (EPFL), Station 3, CH-1015-Lausanne, Switzerland}
\affiliation{Swiss Light Source (SLS), Paul Scherrer Institute (PSI), CH-5232 Villigen PSI, Switzerland}
\author{C.~Piamonteze}
\affiliation{Swiss Light Source (SLS), Paul Scherrer Institute (PSI), CH-5232 Villigen PSI, Switzerland}
\author{C.~P.~Lutz}
\affiliation{IBM Almaden Research Center, 650 Harry Road, San Jose, CA 95120, USA}
\author{R.~M.~Macfarlane}
\affiliation{IBM Almaden Research Center, 650 Harry Road, San Jose, CA 95120, USA}
\author{B.~A.~Jones}
\affiliation{IBM Almaden Research Center, 650 Harry Road, San Jose, CA 95120, USA}
\author{P.~Gambardella}
\affiliation{Department of Materials, ETH Z{\"u}rich H{\"o}nggerbergring 64, CH-8093 Z{\"u}rich, Switzerland}
\author{A.~ J.~Heinrich}
\affiliation{IBM Almaden Research Center, 650 Harry Road, San Jose, CA 95120, USA}
\author{H.~Brune}
\affiliation{Institute of Condensed Matter Physics, {\'E}cole Polytechnique F{\'e}d{\'e}rale de Lausanne (EPFL), Station 3, CH-1015-Lausanne, Switzerland}

\begin{abstract}
We combine density functional theory, x-ray magnetic circular dichroism, multiplet calculations, and scanning tunneling spectroscopy to assess the magnetic properties of Fe atoms adsorbed on a thin layer of MgO(100) on Ag(100). Despite the strong axial field due to the O ligand, the weak cubic field induced by the four-fold coordination to Mg atoms entirely quenches the first order orbital moment. This is in marked contrast to Co, which has an out-of-plane orbital moment of $L_z = \pm 3$ that is protected from mixing in a cubic ligand field. The spin-orbit interaction restores a large fraction of the Fe orbital moment leading a zero-field splitting of $14.0 \pm 0.3$~meV, the largest value reported for surface adsorbed Fe atoms.
\end{abstract}
\maketitle

The magnetic properties of individual atoms adsorbed onto single crystal surfaces are remarkable. Their magnetic anisotropy energies are several orders of magnitude higher than in bulk~\cite{gam03, hir07, mei08, ott08, bru09, don13, don14}. The main cause are the crystal field provided by the surface and the large orbital magnetic moment preserved by the low coordinated adatoms. This moment interacts with the ligand field of the adsorption site and is coupled to the spin moment by spin-orbit coupling (SOC). A system where the adatom exhibits the full gas-phase orbital moment is Co on a few monolayers (ML) of MgO(100) grown epitaxially on Ag(100)~\cite{rau14}. The binding of Co to the underlying O is strong while it is very weak to its next-nearest Mg neighbors. This creates an almost perfect axial crystal field for the Co atom enabling the maximum possible magnetic anisotropy energy for a 3$d$ atom. Moreover, the MgO layer decouples the atoms from the conduction electrons of the underlying metal substrate reducing spin-flip scattering and increasing the magnetic relaxation times.

To understand how individual Fe atoms react to this peculiar adsorption environment, we studied the magnetic properties of Fe/MgO by combining density functional theory (DFT), x-ray magnetic circular dichroism (XMCD), multiplet calculations, and spin excitation spectroscopy (SES) with the scanning tunneling microscope (STM)~\cite{hei04}. We note that Fe ions in bulk MgO have long been considered as a model system for understanding the interplay between crystal field and SOC effects in 3$d$ metals~\cite{ham69, hau10}. This interplay gives rise to a manyfold of low-lying excited states that strongly affects the magnitude of the orbital and spin moments and depends in a subtle way on local symmetry deviations from the ideal cubic environment. The discovery of high tunnel magnetoresistance (TMR)~\cite{mat01, yua04} and perpendicular magnetic anisotropy in Fe/Mg(100)~\cite{kla01, yan11} interfaces have strongly renewed the interest in that system due to technological applications in magnetic recording~\cite{par04, ike10, cub14}. Understanding the delicate balance among crystal field, SOC, and orbital hybridization between Fe and MgO is thus crucial, also to optimize the magnetic properties and downscaling of TMR devices.

We find the lowest-lying zero-magnetic-field excitation at 14~meV. This energy is significantly larger than previous reported values for Fe~\cite{hir07}, but also much smaller than the zero-field splitting of 58~meV reported for Co~\cite{rau14}. This difference between both elements is linked to the orbital symmetry of the magnetic atom. The low-lying orbital states of Fe $L_z = \pm 2$ are mixed in the ligand field of the four next nearest-neighbor atoms (Mg). This fully quenches the orbital moment. In contrast, Co has $L_z = \pm 3$ and thereby is protected from such mixing. The spin-orbit interaction restores a significant fraction of the orbital moment for Fe, while it leaves the ground state orbital and spin moments largely unchanged for Co. The resulting second-order orbital moment observed for Fe is the prevalent situation when atomic spins are incorporated in low-symmetry bonding geometries at surfaces, in bulk, and in molecules~\cite{mcg66, gat06}.

\begin{figure}[t]
\includegraphics[width = 8.5 cm]{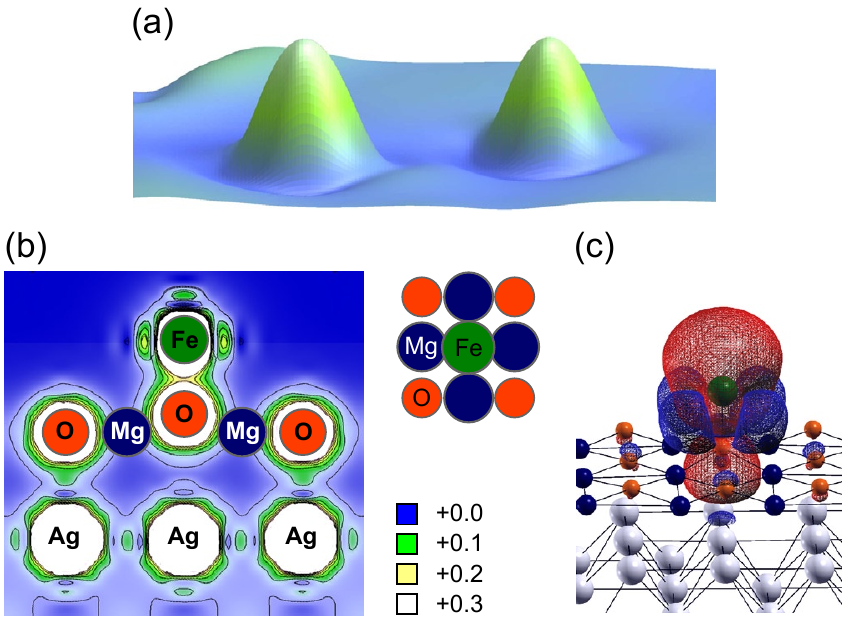}
\caption{(color online). (a) STM image of two Fe atoms on a ML MgO(100) grown on Ag(100) ($4 \ {\rm nm} \times 4$~nm, tunnel current $I_{\rm t} = 5$~pA, tunnel voltage $V_{\rm t} = 100$~mV). (b) Side view of DFT calculated binding geometry and charge density (color scale $1 e / ({\rm au})^3$, Fe green, O red, Mg blue). Middle sketch shows top view ball model of the binding geometry. (c) Oblique view of DFT-calculated valence electron spin density contours (positive spin polarization -- red, negative -- blue).}
\label{SD}
\end{figure}

Figure~\ref{SD}(a) shows an STM image of two individual Fe atoms deposited at $\approx 8$~K on one ML MgO(100) grown on Ag(100)~\cite{sin01, ney04, bau14, sup}. At the Fe coverages of 0.01--0.03~ML (one ML is defined as one Fe atom per MgO(100) unit cell) used in the present study, we observe isolated Fe atoms rather than clusters. We find only one Fe species with an apparent height of 180~pm. The lowest energy adsorption site found in DFT is the one on-top of oxygen~\cite{ney04, fer15, sup}. Figure~\ref{SD}(b) reveals that the O beneath the Fe is displaced by 40~pm upwards from the MgO plane~\cite{sup}. The ground state electron density shows a significant charge transfer between the Fe and the nearest O atom with an overall positive charge of +0.44~$e$ on the Fe~\cite{sup}. Note that Co on the same binding site is nearly charge neutral~\cite{rau14}.

With the generalized gradient approximation (GGA) and on-site Coulomb interactions ($U = 3.2$~eV)~\cite{coc05} for the Fe $d$-states, we find a total spin moment of $3.6 \; \mu_{\rm B}$ on the Fe atom. The majority spin density, red in Fig.~\ref{SD}(c), is mostly axially symmetric and exhibits an induced polarization of the O atom slightly increasing the overall spin moment to $3.7 \; \mu_{\rm B}$. The four-fold symmetry of the binding site shows up strongly in the minority spin density (blue) in contrast to Co, which exhibits nearly perfect axial spin density~\cite{rau14}. The calculated spin of the Fe atom is to very good approximation independent of the MgO thickness. This facilitates the interpretation of synchrotron measurements on samples having several coexisting MgO thicknesses~\cite{sup}.

\begin{figure}[b]
\includegraphics[width = 8.5 cm]{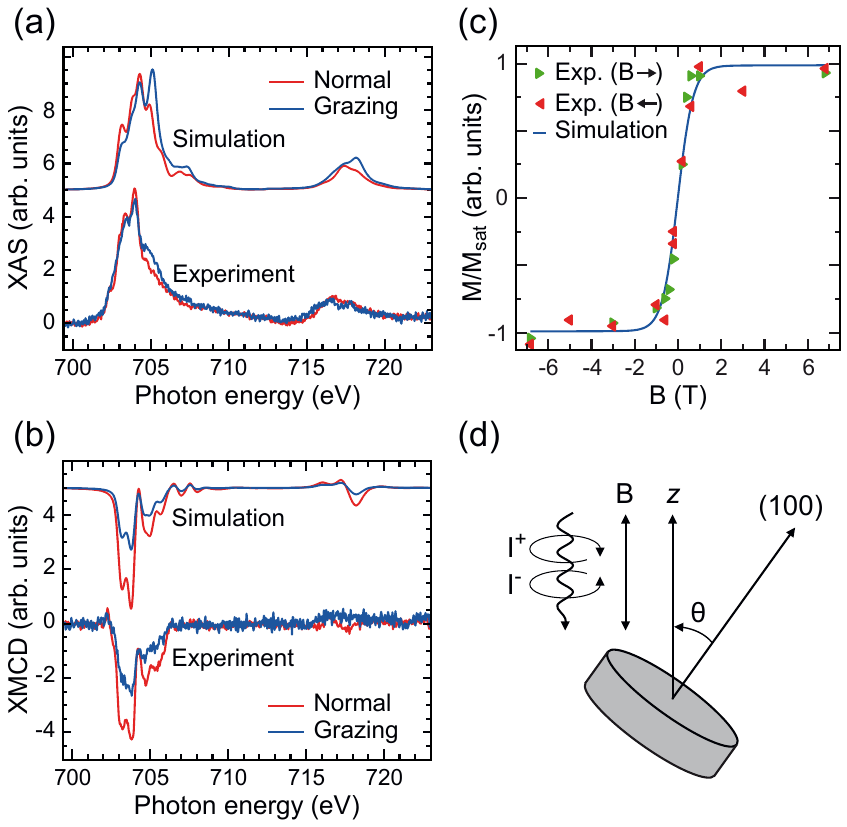}
\caption{(color online). (a) Measured and simulated XAS over the Fe $L_3$ and $L_2$ edges for 0.03~ML Fe on an MgO film on Ag(100) with an average thickness of 3 ML. Red curves are for normal ($\Theta = 0^{\circ}$) and blue for grazing ($\Theta = 60^{\circ}$) incidence of x-ray beam and $B$ field ($T = 2.5$~K, $B = 6.8$~T, total electron yield mode). (b) XMCD spectra for both geometries. (c) Out-of-plane magnetization curve measured by first saturating the sample at 6.8~T (red) and $-6.8$~T (green) and then moving to the respective field value ($T = 2.5$~K). The solid line represents $\langle 2 S_z(B) \rangle + \langle L_z(B) \rangle$ determined by the multiplet fit with a saturation moment of $5.2 \, \mu_{\rm B}$. (d) Sketch of the measurement geometry.}
\label{XMCD}
\end{figure}

X-ray absorption spectra (XAS) and the resulting XMCD signal~\cite{pia12} are shown in Figs.~\ref{XMCD}(a) and (b) (see Ref.~\cite{sup} for experimental details). The Fe $L_3$ and $L_2$ edges exhibit sharp multiplet structure characteristic of an ensemble of single adatoms on identical adsorption sites~\cite{gam02}. The XMCD signal is larger at normal than at grazing incidence, revealing out-of-plane easy magnetization axis. The sum rules~\cite{tho92,car93,che95} yield a large orbital moment of $\langle L_z \rangle = 1.74 \pm 0.11 \, \mu_{\rm B}$ and an effective spin moment of $\langle 2 S_z \rangle + \langle 7 T_z \rangle = 2.46 \pm 0.11 \, \mu_{\rm B}$ (assuming 3.9 $d$ holes, as found in the multiplet calculations below; $T_z$ is the out-of-plane projection of the atomic magnetic dipole moment). The measured out-of-plane magnetization curve is shown in Fig.~\ref{XMCD}(c).

More insight into the magnetic levels and the evolution of their energies is gained from multiplet theory~\cite{gro01}. The calculated spectra shown in Fig.~\ref{XMCD} are in very good agreement with experiment for both incident beam directions. In addition, the experimental magnetization curve is perfectly reproduced by the line showing the out-of-plane projected field-dependent total magnetic moment $\langle 2 S_z(B) \rangle + \langle L_z(B) \rangle$ derived from multiplet calculations. In these calculations we included charge transfer to the O ligand, leading to configuration mixing, the axial ligand field due to the nearest-neighbor O atom ($Ds$), the cubic one due to the four next-nearest neighbor Mg atoms ($Dq$), SOC ($\zeta$), and the external magnetic field ($B$). Best agreement is obtained with a 90~\% $d 6$ $+$ 10~\% $d 7 \, l$ configuration of the Fe atom, where $l$ refers to a ligand hole in the neighboring O atom.

\begin{figure}
\includegraphics[width = 8.5 cm]{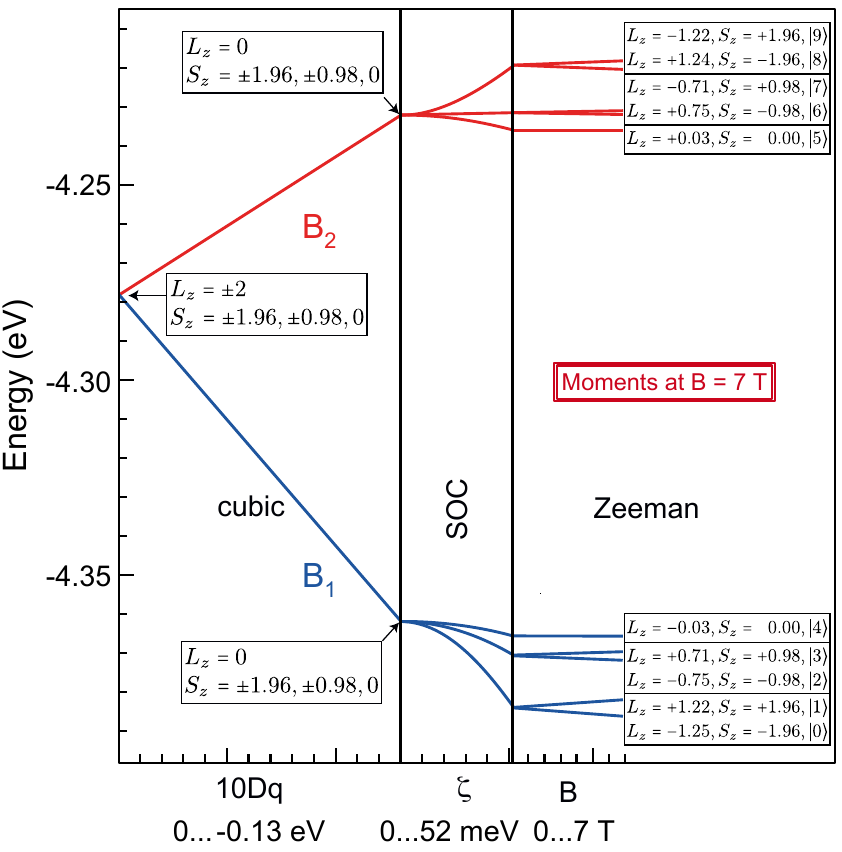}
\caption{(color online). Energy level diagram resulting from the multiplet simulation of the XAS and XMCD spectra (for full diagram see~\cite{sup}). The cubic crystal field quenches the orbital moment and creates two spin quintuplets with $B_1$ (blue) and $B_2$ (red) symmetry. SOC creates an essentially 2-fold degenerate ground state with largely restored orbital moment of $\langle L_z \rangle = \pm 1.25$. The $S_z$ and $L_z$ values in the labels are the respective expectation values, the $\langle \rangle$ signs have been omitted for brevity. The Zeeman energy splits these states into the five states $| 0 \rangle$ -- $| 4 \rangle$. The consecutive states deriving from the $B_2$ manifold are labelled $| 5 \rangle$ -- $| 9 \rangle$.}
\label{multi}
\end{figure}

The configuration mixing results in a 10-fold degenerate ground state. $Ds$ moves this state down in energy, while all other states move up. The resulting ground state $( \langle L_z \rangle = \pm 2 ) \otimes ( \langle S_z \rangle = \pm 1.96$, $\pm 0.98$, and $0 )$ is shown on the left hand side of Fig.~\ref{multi} that illustrates its evolution under the action of $Dq$, $\zeta$, and $B$. In marked contrast to Co, $Dq$ strongly perturbs the lowest multiplet and creates two spin quintuplets with fully quenched orbital moments, $( \langle L_z \rangle = 0 ) \otimes ( \langle S_z \rangle = \pm 1.96$, $\pm 0.98$, and $0 )$. The one with $B_1$ symmetry is the ground state and drawn in blue. SOC splits it into three energy levels, where the lowest restores more than half of the free-atom orbital moment by coupling the two lowest orbital levels in a second-order perturbation~\cite{mcg66}. At zero magnetic field, the new ground states $\langle L_z \rangle = \pm 1.25$ and $\langle S_z \rangle = \pm 1.96$ are to a very good approximation two-fold degenerate (ignoring energy differences of a few neV). The magnetic field lifts the remaining degeneracy of the five states labelled $| 0 \rangle$ -- $| 4 \rangle$. The excited spin quintuplet (red) has $B_2$ symmetry, lies $\approx 100$~meV higher in energy, and has its orbital and spin magnetic moments anti-aligned. This results in smaller total magnetic moments and hence a smaller Zeeman splitting.

\begin{figure}
\includegraphics[width = 8.5 cm]{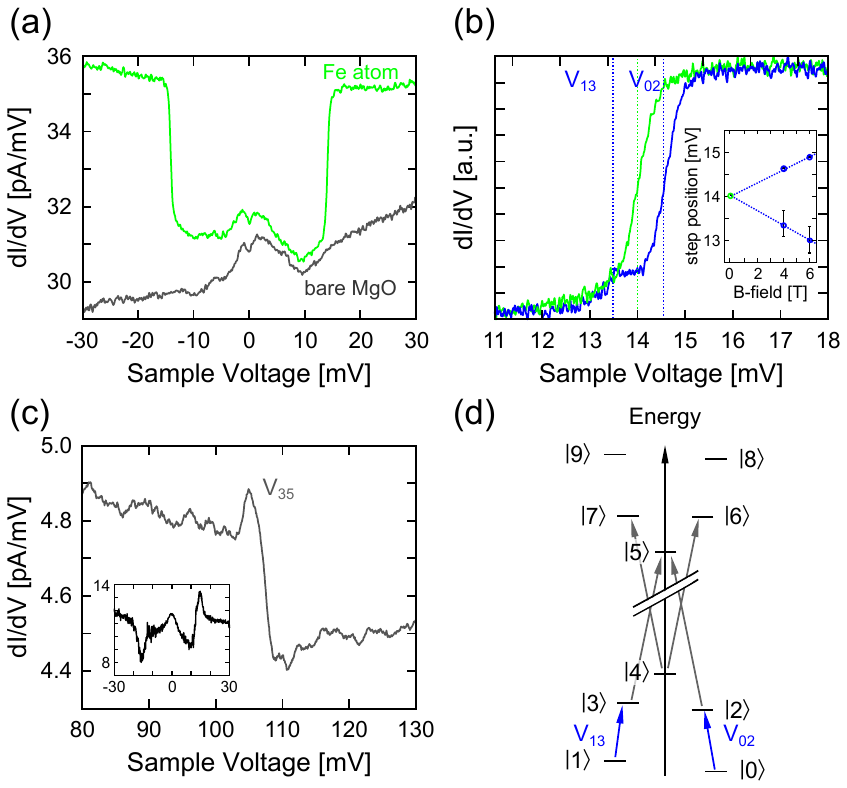}
\caption{(color online). (a) STM differential conductance ($dI/dV$) spectrum on Fe/1~ML MgO(100)/Ag(100), $dI/dV$ on bare MgO is shown for comparison ($T = 0.6$~K, $B = 0$~T, $V_{\rm t}$ modulation with rms amplitude $V_{\rm mod} = 150 \; \mu$V and frequency $f = 806$~Hz, set point before opening the feedback loop $I_{\rm t} = 1$~nA, $V_{\rm t} = 30$~mV). (b) Positive conductance step at out-of-plane fields of 0~T (green) and 4~T  (blue). Inset: Field splitting of conductance step energies. (c) SES feature corresponding to the superposition of $V_{35}$, $V_{25}$, $V_{47}$, and $V_{46}$, measured with a spin-polarized tip ($T = 1.2$~K, $B = 2$~T, $V_{\rm mod} = 1.5$~mV, $I_{\rm t} = 1$~nA, $V_{\rm t} = 100$). Inset: spin-polarized spectrum in the same energy window as (a). (d) Sketch of the magnetic states and the allowed SES excitations.}
\label{SES}
\end{figure}

Our STM-SES measurements on individual Fe atoms determine the energy splitting of the lowest lying magnetic states with high precision, and thereby complement XMCD. Figure~\ref{SES}(a) displays clear conductance steps with 15~\% amplitude located at $\pm 14.0 \pm 0.3$~mV (the error bar refers to variations between atoms at different locations of the MgO film). The magnetic nature of the underlying inelastic excitations is demonstrated by the splitting of the excitation energy in an out-of-plane magnetic field shown in Fig.~\ref{SES}(b). For in-plane fields the splitting is absent~\cite{sup}, indicative of an out-of-plane easy axis. The observed zero-field splitting of 14~meV is more than twice the largest values seen for individual Fe atoms adsorbed on other surfaces~\cite{hir07} and our magnetic anisotropy energy~\cite{mae} approaches that reported for Fe atoms in linear molecules~\cite{zad13}.

Connection between the transitions excited in SES and the states derived from the multiplet calculations is established by the level diagram shown in Fig.~\ref{SES}(d). The horizontal position of the states $| 0 \rangle$ -- $| 9 \rangle$ indicates their out-of-plane projected magnetic moments. The blue arrows are the first two excitations possible for tunnel electrons, $| 0 \rangle \rightarrow | 2 \rangle$ and $| 1 \rangle \rightarrow | 3 \rangle$, with the corresponding tunnel voltages labeled $V_{02}$ and $V_{13}$. These are the excitations giving rise to the conductance steps in Figs.~\ref{SES}(a) and (b). At zero field, $V_{02} = V_{13} = 14.0$~mV. This is in excellent agreement with the level separation of 13~meV found in the multiplet calculations.

As expected from the Zeeman splitting in Fig.~\ref{multi}, V$_{02}$ and V$_{13}$ shift up and down symmetrically in an external out-of-plane field, see Fig.~\ref{SES}(b). As seen from the inset, the splitting is linear in field and amounts to $1.9 \pm 0.3$~mV at 6~T. The amplitude of $V_{13}$ is significantly smaller than that of $V_{02}$ because the corresponding transition starts from the excited state $| 1 \rangle$. That $V_{13}$ is visible at all implies that the tunnel current drives the occupation of the $| 1 \rangle$ state, and that its lifetime is longer than the mean time between tunneling electrons of the order of 1~ns~\cite{lot10a}. From the slopes in the inset of Fig.~\ref{SES}(b) we derive an effective $g^*$-value of $2.57 \pm 0.06$ in the $z$-direction. This is significantly above the free-electron value of $g_e = 2$ and thus confirms the presence of a large unquenched orbital magnetic moment for Fe on MgO~\cite{chi11, suh11}. From XMCD and the multiplet calculations we derive $g^*$ as the difference of the out-of-plane projected magnetic moments between states $| 0 \rangle$ and $| 2 \rangle$, $g^* = \Delta \langle L_z \rangle + 2 \Delta \langle S_z \rangle = 2.46$, again in very good agreement with SES.

The magnitude of the zero-field splitting and of the orbital moment strongly depend on the cubic ligand field. However, the simulated spectra are not very sensitive to small changes of $Dq$. This illustrates the strength in combining SES and XMCD. The first measures the zero-field splitting directly and very precisely, while the second identifies the ground and excited states with their spin and orbital magnetic moments, as well as their angular anisotropies.

The strong sensitivity to the cubic ligand field shows that Fe exhibits a fundamentally different magnetic behavior on the MgO surface compared to Co on the same substrate~\cite{rau14}. The key difference is that the cubic field is able to quench the Fe orbital moment because the four-fold symmetry can mix the $L_z = \pm 2$ components, resulting in $\langle L_z \rangle = 0$ after the ligand field contribution. The $L_z = \pm 3$ components of Co are protected from such mixing, leading to a linear response to SOC~\cite{rau14}. Fe represents the more common case~\cite{abr86}, where SOC restores some of the otherwise quenched orbital moment, due to mixing with the next-higher orbital state.

Spin-polarized (SP) STM tips reveal an additional conductance step at $\sim 105$~mV, see Fig.~\ref{SES}(c). This step is absent for non-polarized tips. Therefore, we assign it to an electron driven occupation change of the magnetic states at the respective threshold energy (spin pumping). Evidently, this part of the conductance change in spin polarized spectra depends on the set-point current, since the tunneling electrons must arrive frequently enough to probe the excited states before they decay~\cite{lot10a}. At the $V_{02}$ and $V_{13}$ step energy, inelastic spin excitations and spin pumping are superimposed creating the jagged edges shown in the inset of Fig.~\ref{SES}(c).

In order to identify which states contribute to the 105~mV step, we note that the transition rate from the initial to a final state follows Fermi's golden rule and is proportional to the interaction potential of the atom's spin with the spin of the tunneling electron~\cite{hir07, lor09, lot10b, sup}. This model implicitly requires that the orbital moments remain constant, while the tunneling electrons can exchange spin angular momentum with the atom according to the usual selection rule $\Delta S_z = \pm 1, \, 0$. Note that the selection rule of the orbital moments is fulfilled if part of the initial orbital wave function overlaps with the final orbital wave function. This is the case for all lowest 10 states of Fe, as they all have components of $L_z = \pm 2$, despite the changing expectation value $\langle L_z \rangle$. Using these selection rules, we can assign the $\sim 105$~mV step to the four transitions $V_{35}$, $V_{25}$, $V_{46}$, and $V_{47}$ that are very close in energy~\cite{v06}. This again implies lifetimes $\tau > 1$~ns of one or more of the initial states $| 2 \rangle$, $| 3 \rangle$, and $| 4 \rangle$. The multiplet calculations estimate $V_{35} \sim 134$~meV and predict a transition probability one order of magnitude smaller than for the $V_{02}$ transition~\cite{sup}. The small intensity explains the absence of this step in non-polarized $dI/dV$ spectra. The energy is in reasonable agreement with the one directly probed by STM.

Transitions between states belonging to different spin multiplets have been observed in spin chains ({\it e.g.} singlet to doublet and triplet states)~\cite{hir06, lot10a} and molecular magnets~\cite{kah12}. Unique to Fe on MgO, the excitation from lower to upper spin multiplet involves a transition from aligned (lower multiplet) to anti-aligned (upper multiplet) orbital and spin moments.

The present study reveals how the perpendicular magnetic anisotropy and enhanced orbital moment of Fe atoms on MgO arise from the interplay of SOC and the low-symmetry ligand field at O sites. Despite the similar chemical environment, we find important differences between Fe and Co adsorbed onto MgO thin films. For both elements, the axial ligand field of the oxygen fully preserves the gas phase orbital moments of $L = 2$, respectively, $L = 3$. However, the cubic ligand field created by the much weaker interaction with the next-nearest neighbors (Mg) acts very differently on both elements. For Fe it quenches the orbital moment due to mixing of the $L_z = \pm 2$ components of the orbital wave function, while the $L_z = \pm 3$ components of Co are protected from that mixing by symmetry. In the case of Fe, SOC restores a very large fraction of the orbital moment of $\langle L_z \rangle = 1.25 \, \mu_{\rm B}$, leading to a high zero-field splitting of 14~meV unprecedented for Fe adatoms. The magnetic levels derived from XMCD and multiplet calculations agree extremely well with the excitation energies, their field splitting, and excitation amplitudes observed in STM-SES. Remarkably, spin-polarized STM can also probe excitations to higher spin multiplets.

\begin{acknowledgements}
We acknowledge funding from the Swiss National Science Foundation, the Competence Centre for Materials Science and Technology, the COST action MP0903, and the Office of Naval Research. We thank Bruce Melior for expert technical assistance. We used computational resources of  NERSC and  IBM Research Almaden in house Blue Gene P for our DFT calculations and data visualization.
\end{acknowledgements}

\bibliographystyle{apsrev4-1}
%\bibliography{ms_7}

\begin{thebibliography}{53}%
\makeatletter
\providecommand \@ifxundefined [1]{%
 \@ifx{#1\undefined}
}%
\providecommand \@ifnum [1]{%
 \ifnum #1\expandafter \@firstoftwo
 \else \expandafter \@secondoftwo
 \fi
}%
\providecommand \@ifx [1]{%
 \ifx #1\expandafter \@firstoftwo
 \else \expandafter \@secondoftwo
 \fi
}%
\providecommand \natexlab [1]{#1}%
\providecommand \enquote  [1]{``#1''}%
\providecommand \bibnamefont  [1]{#1}%
\providecommand \bibfnamefont [1]{#1}%
\providecommand \citenamefont [1]{#1}%
\providecommand \href@noop [0]{\@secondoftwo}%
\providecommand \href [0]{\begingroup \@sanitize@url \@href}%
\providecommand \@href[1]{\@@startlink{#1}\@@href}%
\providecommand \@@href[1]{\endgroup#1\@@endlink}%
\providecommand \@sanitize@url [0]{\catcode `\\12\catcode `\$12\catcode
  `\&12\catcode `\#12\catcode `\^12\catcode `\_12\catcode `\%12\relax}%
\providecommand \@@startlink[1]{}%
\providecommand \@@endlink[0]{}%
\providecommand \url  [0]{\begingroup\@sanitize@url \@url }%
\providecommand \@url [1]{\endgroup\@href {#1}{\urlprefix }}%
\providecommand \urlprefix  [0]{URL }%
\providecommand \Eprint [0]{\href }%
\providecommand \doibase [0]{http://dx.doi.org/}%
\providecommand \selectlanguage [0]{\@gobble}%
\providecommand \bibinfo  [0]{\@secondoftwo}%
\providecommand \bibfield  [0]{\@secondoftwo}%
\providecommand \translation [1]{[#1]}%
\providecommand \BibitemOpen [0]{}%
\providecommand \bibitemStop [0]{}%
\providecommand \bibitemNoStop [0]{.\EOS\space}%
\providecommand \EOS [0]{\spacefactor3000\relax}%
\providecommand \BibitemShut  [1]{\csname bibitem#1\endcsname}%
\let\auto@bib@innerbib\@empty
%</preamble>
\bibitem [{\citenamefont {Gambardella}\ \emph {et~al.}(2003)\citenamefont
  {Gambardella}, \citenamefont {Rusponi}, \citenamefont {Veronese},
  \citenamefont {Dhesi}, \citenamefont {Grazioli}, \citenamefont {Dallmeyer},
  \citenamefont {Cabria}, \citenamefont {Zeller}, \citenamefont {Dederichs},
  \citenamefont {Kern}, \citenamefont {Carbone},\ and\ \citenamefont
  {Brune}}]{gam03}%
  \BibitemOpen
  \bibfield  {author} {\bibinfo {author} {\bibfnamefont {P.}~\bibnamefont
  {Gambardella}}, \bibinfo {author} {\bibfnamefont {S.}~\bibnamefont
  {Rusponi}}, \bibinfo {author} {\bibfnamefont {M.}~\bibnamefont {Veronese}},
  \bibinfo {author} {\bibfnamefont {S.~S.}\ \bibnamefont {Dhesi}}, \bibinfo
  {author} {\bibfnamefont {C.}~\bibnamefont {Grazioli}}, \bibinfo {author}
  {\bibfnamefont {A.}~\bibnamefont {Dallmeyer}}, \bibinfo {author}
  {\bibfnamefont {I.}~\bibnamefont {Cabria}}, \bibinfo {author} {\bibfnamefont
  {R.}~\bibnamefont {Zeller}}, \bibinfo {author} {\bibfnamefont {P.~H.}\
  \bibnamefont {Dederichs}}, \bibinfo {author} {\bibfnamefont {K.}~\bibnamefont
  {Kern}}, \bibinfo {author} {\bibfnamefont {C.}~\bibnamefont {Carbone}}, \
  and\ \bibinfo {author} {\bibfnamefont {H.}~\bibnamefont {Brune}},\
  }\href@noop {} {\bibfield  {journal} {\bibinfo  {journal} {Science}\ }\textbf
  {\bibinfo {volume} {300}},\ \bibinfo {pages} {1130} (\bibinfo {year}
  {2003})}\BibitemShut {NoStop}%
\bibitem [{\citenamefont {Hirjibehedin}\ \emph {et~al.}(2007)\citenamefont
  {Hirjibehedin}, \citenamefont {Lin}, \citenamefont {Otte}, \citenamefont
  {Ternes}, \citenamefont {Lutz}, \citenamefont {Jones},\ and\ \citenamefont
  {Heinrich}}]{hir07}%
  \BibitemOpen
  \bibfield  {author} {\bibinfo {author} {\bibfnamefont {C.~F.}\ \bibnamefont
  {Hirjibehedin}}, \bibinfo {author} {\bibfnamefont {C.~Y.}\ \bibnamefont
  {Lin}}, \bibinfo {author} {\bibfnamefont {A.~F.}\ \bibnamefont {Otte}},
  \bibinfo {author} {\bibfnamefont {M.}~\bibnamefont {Ternes}}, \bibinfo
  {author} {\bibfnamefont {C.~P.}\ \bibnamefont {Lutz}}, \bibinfo {author}
  {\bibfnamefont {B.~A.}\ \bibnamefont {Jones}}, \ and\ \bibinfo {author}
  {\bibfnamefont {A.~J.}\ \bibnamefont {Heinrich}},\ }\href@noop {} {\bibfield
  {journal} {\bibinfo  {journal} {Science}\ }\textbf {\bibinfo {volume}
  {317}},\ \bibinfo {pages} {1199} (\bibinfo {year} {2007})}\BibitemShut
  {NoStop}%
\bibitem [{\citenamefont {Meier}\ \emph {et~al.}(2008)\citenamefont {Meier},
  \citenamefont {Zhou}, \citenamefont {Wiebe},\ and\ \citenamefont
  {Wiesendanger}}]{mei08}%
  \BibitemOpen
  \bibfield  {author} {\bibinfo {author} {\bibfnamefont {F.}~\bibnamefont
  {Meier}}, \bibinfo {author} {\bibfnamefont {L.}~\bibnamefont {Zhou}},
  \bibinfo {author} {\bibfnamefont {J.}~\bibnamefont {Wiebe}}, \ and\ \bibinfo
  {author} {\bibfnamefont {R.}~\bibnamefont {Wiesendanger}},\ }\href@noop {}
  {\bibfield  {journal} {\bibinfo  {journal} {Science}\ }\textbf {\bibinfo
  {volume} {320}},\ \bibinfo {pages} {82} (\bibinfo {year} {2008})}\BibitemShut
  {NoStop}%
\bibitem [{\citenamefont {Otte}\ \emph {et~al.}(2008)\citenamefont {Otte},
  \citenamefont {Ternes}, \citenamefont {von Bergmann}, \citenamefont {Loth},
  \citenamefont {Brune}, \citenamefont {Lutz}, \citenamefont {Hirjibehedin},\
  and\ \citenamefont {Heinrich}}]{ott08}%
  \BibitemOpen
  \bibfield  {author} {\bibinfo {author} {\bibfnamefont {A.~F.}\ \bibnamefont
  {Otte}}, \bibinfo {author} {\bibfnamefont {M.}~\bibnamefont {Ternes}},
  \bibinfo {author} {\bibfnamefont {K.}~\bibnamefont {von Bergmann}}, \bibinfo
  {author} {\bibfnamefont {S.}~\bibnamefont {Loth}}, \bibinfo {author}
  {\bibfnamefont {H.}~\bibnamefont {Brune}}, \bibinfo {author} {\bibfnamefont
  {C.~P.}\ \bibnamefont {Lutz}}, \bibinfo {author} {\bibfnamefont {C.~F.}\
  \bibnamefont {Hirjibehedin}}, \ and\ \bibinfo {author} {\bibfnamefont
  {A.~J.}\ \bibnamefont {Heinrich}},\ }\href@noop {} {\bibfield  {journal}
  {\bibinfo  {journal} {Nat. Phys.}\ }\textbf {\bibinfo {volume} {4}},\
  \bibinfo {pages} {847} (\bibinfo {year} {2008})}\BibitemShut {NoStop}%
\bibitem [{\citenamefont {Brune}\ and\ \citenamefont
  {Gambardella}(2009)}]{bru09}%
  \BibitemOpen
  \bibfield  {author} {\bibinfo {author} {\bibfnamefont {H.}~\bibnamefont
  {Brune}}\ and\ \bibinfo {author} {\bibfnamefont {P.}~\bibnamefont
  {Gambardella}},\ }\href@noop {} {\bibfield  {journal} {\bibinfo  {journal}
  {Surf. Sci.}\ }\textbf {\bibinfo {volume} {603}},\ \bibinfo {pages} {1812}
  (\bibinfo {year} {2009})}\BibitemShut {NoStop}%
\bibitem [{\citenamefont {Donati}\ \emph {et~al.}(2013)\citenamefont {Donati},
  \citenamefont {Dubout}, \citenamefont {Autès}, \citenamefont {Patthey},
  \citenamefont {Calleja}, \citenamefont {Gambardella}, \citenamefont
  {Yazyev},\ and\ \citenamefont {Brune}}]{don13}%
  \BibitemOpen
  \bibfield  {author} {\bibinfo {author} {\bibfnamefont {F.}~\bibnamefont
  {Donati}}, \bibinfo {author} {\bibfnamefont {Q.}~\bibnamefont {Dubout}},
  \bibinfo {author} {\bibfnamefont {G.}~\bibnamefont {Autès}}, \bibinfo
  {author} {\bibfnamefont {F.}~\bibnamefont {Patthey}}, \bibinfo {author}
  {\bibfnamefont {F.}~\bibnamefont {Calleja}}, \bibinfo {author} {\bibfnamefont
  {P.}~\bibnamefont {Gambardella}}, \bibinfo {author} {\bibfnamefont {O.~V.}\
  \bibnamefont {Yazyev}}, \ and\ \bibinfo {author} {\bibfnamefont
  {H.}~\bibnamefont {Brune}},\ }\href@noop {} {\bibfield  {journal} {\bibinfo
  {journal} {Phys. Rev. Lett.}\ }\textbf {\bibinfo {volume} {111}},\ \bibinfo
  {pages} {236801} (\bibinfo {year} {2013})}\BibitemShut {NoStop}%
\bibitem [{\citenamefont {Donati}\ \emph {et~al.}(2014)\citenamefont {Donati},
  \citenamefont {Gragnaniello}, \citenamefont {Cavallin}, \citenamefont
  {Natterer}, \citenamefont {Dubout}, \citenamefont {Pivetta}, \citenamefont
  {Patthey}, \citenamefont {Dreiser}, \citenamefont {Piamonteze}, \citenamefont
  {Rusponi},\ and\ \citenamefont {Brune}}]{don14}%
  \BibitemOpen
  \bibfield  {author} {\bibinfo {author} {\bibfnamefont {F.}~\bibnamefont
  {Donati}}, \bibinfo {author} {\bibfnamefont {L.}~\bibnamefont
  {Gragnaniello}}, \bibinfo {author} {\bibfnamefont {A.}~\bibnamefont
  {Cavallin}}, \bibinfo {author} {\bibfnamefont {F.~D.}\ \bibnamefont
  {Natterer}}, \bibinfo {author} {\bibfnamefont {Q.}~\bibnamefont {Dubout}},
  \bibinfo {author} {\bibfnamefont {M.}~\bibnamefont {Pivetta}}, \bibinfo
  {author} {\bibfnamefont {F.}~\bibnamefont {Patthey}}, \bibinfo {author}
  {\bibfnamefont {J.}~\bibnamefont {Dreiser}}, \bibinfo {author} {\bibfnamefont
  {C.}~\bibnamefont {Piamonteze}}, \bibinfo {author} {\bibfnamefont
  {S.}~\bibnamefont {Rusponi}}, \ and\ \bibinfo {author} {\bibfnamefont
  {H.}~\bibnamefont {Brune}},\ }\href@noop {} {\bibfield  {journal} {\bibinfo
  {journal} {Phys. Rev. Lett.}\ }\textbf {\bibinfo {volume} {113}},\ \bibinfo
  {pages} {177201} (\bibinfo {year} {2014})}\BibitemShut {NoStop}%
\bibitem [{\citenamefont {Rau}\ \emph {et~al.}(2014)\citenamefont {Rau},
  \citenamefont {Baumann}, \citenamefont {Rusponi}, \citenamefont {Donati},
  \citenamefont {Stepanow}, \citenamefont {Gragnaniello}, \citenamefont
  {Dreiser}, \citenamefont {Piamonteze}, \citenamefont {Nolting}, \citenamefont
  {Gangopadhyay}, \citenamefont {Albertini}, \citenamefont {Macfarlane},
  \citenamefont {Lutz}, \citenamefont {Jones}, \citenamefont {Gambardella},
  \citenamefont {Heinrich},\ and\ \citenamefont {Brune}}]{rau14}%
  \BibitemOpen
  \bibfield  {author} {\bibinfo {author} {\bibfnamefont {I.~G.}\ \bibnamefont
  {Rau}}, \bibinfo {author} {\bibfnamefont {S.}~\bibnamefont {Baumann}},
  \bibinfo {author} {\bibfnamefont {S.}~\bibnamefont {Rusponi}}, \bibinfo
  {author} {\bibfnamefont {F.}~\bibnamefont {Donati}}, \bibinfo {author}
  {\bibfnamefont {S.}~\bibnamefont {Stepanow}}, \bibinfo {author}
  {\bibfnamefont {L.}~\bibnamefont {Gragnaniello}}, \bibinfo {author}
  {\bibfnamefont {J.}~\bibnamefont {Dreiser}}, \bibinfo {author} {\bibfnamefont
  {C.}~\bibnamefont {Piamonteze}}, \bibinfo {author} {\bibfnamefont
  {F.}~\bibnamefont {Nolting}}, \bibinfo {author} {\bibfnamefont
  {S.}~\bibnamefont {Gangopadhyay}}, \bibinfo {author} {\bibfnamefont {O.~R.}\
  \bibnamefont {Albertini}}, \bibinfo {author} {\bibfnamefont {R.}~\bibnamefont
  {Macfarlane}}, \bibinfo {author} {\bibfnamefont {C.~P.}\ \bibnamefont
  {Lutz}}, \bibinfo {author} {\bibfnamefont {B.~A.}\ \bibnamefont {Jones}},
  \bibinfo {author} {\bibfnamefont {P.}~\bibnamefont {Gambardella}}, \bibinfo
  {author} {\bibfnamefont {A.~J.}\ \bibnamefont {Heinrich}}, \ and\ \bibinfo
  {author} {\bibfnamefont {H.}~\bibnamefont {Brune}},\ }\href@noop {}
  {\bibfield  {journal} {\bibinfo  {journal} {Science}\ }\textbf {\bibinfo
  {volume} {344}},\ \bibinfo {pages} {988} (\bibinfo {year}
  {2014})}\BibitemShut {NoStop}%
\bibitem [{\citenamefont {Heinrich}\ \emph {et~al.}(2004)\citenamefont
  {Heinrich}, \citenamefont {Gupta}, \citenamefont {Lutz},\ and\ \citenamefont
  {Eigler}}]{hei04}%
  \BibitemOpen
  \bibfield  {author} {\bibinfo {author} {\bibfnamefont {A.~J.}\ \bibnamefont
  {Heinrich}}, \bibinfo {author} {\bibfnamefont {J.~A.}\ \bibnamefont {Gupta}},
  \bibinfo {author} {\bibfnamefont {C.~P.}\ \bibnamefont {Lutz}}, \ and\
  \bibinfo {author} {\bibfnamefont {D.~M.}\ \bibnamefont {Eigler}},\
  }\href@noop {} {\bibfield  {journal} {\bibinfo  {journal} {Science}\ }\textbf
  {\bibinfo {volume} {306}},\ \bibinfo {pages} {466} (\bibinfo {year}
  {2004})}\BibitemShut {NoStop}%
\bibitem [{\citenamefont {Ham}\ \emph {et~al.}(1969)\citenamefont {Ham},
  \citenamefont {Schwarz},\ and\ \citenamefont {O'Brien}}]{ham69}%
  \BibitemOpen
  \bibfield  {author} {\bibinfo {author} {\bibfnamefont {F.~S.}\ \bibnamefont
  {Ham}}, \bibinfo {author} {\bibfnamefont {W.~M.}\ \bibnamefont {Schwarz}}, \
  and\ \bibinfo {author} {\bibfnamefont {M.~C.~M.}\ \bibnamefont {O'Brien}},\
  }\href@noop {} {\bibfield  {journal} {\bibinfo  {journal} {Phys. Rev.}\
  }\textbf {\bibinfo {volume} {185}},\ \bibinfo {pages} {548} (\bibinfo {year}
  {1969})}\BibitemShut {NoStop}%
\bibitem [{\citenamefont {Haupricht}\ \emph {et~al.}(2010)\citenamefont
  {Haupricht}, \citenamefont {Sutarto}, \citenamefont {Haverkort},
  \citenamefont {Ott}, \citenamefont {Tanaka}, \citenamefont {Hsieh},
  \citenamefont {Lin}, \citenamefont {Chen}, \citenamefont {Hu},\ and\
  \citenamefont {Tjeng}}]{hau10}%
  \BibitemOpen
  \bibfield  {author} {\bibinfo {author} {\bibfnamefont {T.}~\bibnamefont
  {Haupricht}}, \bibinfo {author} {\bibfnamefont {R.}~\bibnamefont {Sutarto}},
  \bibinfo {author} {\bibfnamefont {M.~W.}\ \bibnamefont {Haverkort}}, \bibinfo
  {author} {\bibfnamefont {H.}~\bibnamefont {Ott}}, \bibinfo {author}
  {\bibfnamefont {A.}~\bibnamefont {Tanaka}}, \bibinfo {author} {\bibfnamefont
  {H.~H.}\ \bibnamefont {Hsieh}}, \bibinfo {author} {\bibfnamefont {H.~J.}\
  \bibnamefont {Lin}}, \bibinfo {author} {\bibfnamefont {C.~T.}\ \bibnamefont
  {Chen}}, \bibinfo {author} {\bibfnamefont {Z.}~\bibnamefont {Hu}}, \ and\
  \bibinfo {author} {\bibfnamefont {L.~H.}\ \bibnamefont {Tjeng}},\ }\href@noop
  {} {\bibfield  {journal} {\bibinfo  {journal} {Phys. Rev. B}\ }\textbf
  {\bibinfo {volume} {82}},\ \bibinfo {pages} {035120} (\bibinfo {year}
  {2010})}\BibitemShut {NoStop}%
\bibitem [{\citenamefont {Mathon}\ and\ \citenamefont {Umerski}(2001)}]{mat01}%
  \BibitemOpen
  \bibfield  {author} {\bibinfo {author} {\bibfnamefont {J.}~\bibnamefont
  {Mathon}}\ and\ \bibinfo {author} {\bibfnamefont {A.}~\bibnamefont
  {Umerski}},\ }\href@noop {} {\bibfield  {journal} {\bibinfo  {journal} {Phys.
  Rev. B}\ }\textbf {\bibinfo {volume} {63}},\ \bibinfo {pages} {220403}
  (\bibinfo {year} {2001})}\BibitemShut {NoStop}%
\bibitem [{\citenamefont {Yuasa}\ \emph {et~al.}(2004)\citenamefont {Yuasa},
  \citenamefont {Nagahama}, \citenamefont {Fukushima}, \citenamefont {Suzuki},\
  and\ \citenamefont {Ando}}]{yua04}%
  \BibitemOpen
  \bibfield  {author} {\bibinfo {author} {\bibfnamefont {S.}~\bibnamefont
  {Yuasa}}, \bibinfo {author} {\bibfnamefont {T.}~\bibnamefont {Nagahama}},
  \bibinfo {author} {\bibfnamefont {A.}~\bibnamefont {Fukushima}}, \bibinfo
  {author} {\bibfnamefont {Y.}~\bibnamefont {Suzuki}}, \ and\ \bibinfo {author}
  {\bibfnamefont {K.}~\bibnamefont {Ando}},\ }\href@noop {} {\bibfield
  {journal} {\bibinfo  {journal} {Nat. Mater.}\ }\textbf {\bibinfo {volume}
  {3}},\ \bibinfo {pages} {868} (\bibinfo {year} {2004})}\BibitemShut {NoStop}%
\bibitem [{\citenamefont {Klaua}\ \emph {et~al.}(2001)\citenamefont {Klaua},
  \citenamefont {Ullmann}, \citenamefont {Barthel}, \citenamefont {Wulfhekel},
  \citenamefont {Kirschner}, \citenamefont {Urban}, \citenamefont {Monchesky},
  \citenamefont {Enders}, \citenamefont {Cochran},\ and\ \citenamefont
  {Heinrich}}]{kla01}%
  \BibitemOpen
  \bibfield  {author} {\bibinfo {author} {\bibfnamefont {M.}~\bibnamefont
  {Klaua}}, \bibinfo {author} {\bibfnamefont {D.}~\bibnamefont {Ullmann}},
  \bibinfo {author} {\bibfnamefont {J.}~\bibnamefont {Barthel}}, \bibinfo
  {author} {\bibfnamefont {W.}~\bibnamefont {Wulfhekel}}, \bibinfo {author}
  {\bibfnamefont {J.}~\bibnamefont {Kirschner}}, \bibinfo {author}
  {\bibfnamefont {R.}~\bibnamefont {Urban}}, \bibinfo {author} {\bibfnamefont
  {T.~L.}\ \bibnamefont {Monchesky}}, \bibinfo {author} {\bibfnamefont
  {A.}~\bibnamefont {Enders}}, \bibinfo {author} {\bibfnamefont {J.~F.}\
  \bibnamefont {Cochran}}, \ and\ \bibinfo {author} {\bibfnamefont
  {B.}~\bibnamefont {Heinrich}},\ }\href@noop {} {\bibfield  {journal}
  {\bibinfo  {journal} {Phys. Rev. B}\ }\textbf {\bibinfo {volume} {64}},\
  \bibinfo {pages} {134411} (\bibinfo {year} {2001})}\BibitemShut {NoStop}%
\bibitem [{\citenamefont {Yang}\ \emph {et~al.}(2011)\citenamefont {Yang},
  \citenamefont {Chshiev}, \citenamefont {Dieny}, \citenamefont {Lee},
  \citenamefont {Manchon},\ and\ \citenamefont {Shin}}]{yan11}%
  \BibitemOpen
  \bibfield  {author} {\bibinfo {author} {\bibfnamefont {H.~X.}\ \bibnamefont
  {Yang}}, \bibinfo {author} {\bibfnamefont {M.}~\bibnamefont {Chshiev}},
  \bibinfo {author} {\bibfnamefont {B.}~\bibnamefont {Dieny}}, \bibinfo
  {author} {\bibfnamefont {J.~H.}\ \bibnamefont {Lee}}, \bibinfo {author}
  {\bibfnamefont {A.}~\bibnamefont {Manchon}}, \ and\ \bibinfo {author}
  {\bibfnamefont {K.~H.}\ \bibnamefont {Shin}},\ }\href@noop {} {\bibfield
  {journal} {\bibinfo  {journal} {Phys. Rev. B}\ }\textbf {\bibinfo {volume}
  {84}},\ \bibinfo {pages} {054401} (\bibinfo {year} {2011})}\BibitemShut
  {NoStop}%
\bibitem [{\citenamefont {Parkin}\ \emph {et~al.}(2004)\citenamefont {Parkin},
  \citenamefont {Kaiser}, \citenamefont {Panchula}, \citenamefont {Rice},
  \citenamefont {Hughes}, \citenamefont {Samant},\ and\ \citenamefont
  {Yang}}]{par04}%
  \BibitemOpen
  \bibfield  {author} {\bibinfo {author} {\bibfnamefont {S.~S.~P.}\
  \bibnamefont {Parkin}}, \bibinfo {author} {\bibfnamefont {C.}~\bibnamefont
  {Kaiser}}, \bibinfo {author} {\bibfnamefont {A.}~\bibnamefont {Panchula}},
  \bibinfo {author} {\bibfnamefont {P.~M.}\ \bibnamefont {Rice}}, \bibinfo
  {author} {\bibfnamefont {B.}~\bibnamefont {Hughes}}, \bibinfo {author}
  {\bibfnamefont {M.}~\bibnamefont {Samant}}, \ and\ \bibinfo {author}
  {\bibfnamefont {S.~H.}\ \bibnamefont {Yang}},\ }\href@noop {} {\bibfield
  {journal} {\bibinfo  {journal} {Nat. Mater.}\ }\textbf {\bibinfo {volume}
  {3}},\ \bibinfo {pages} {862} (\bibinfo {year} {2004})}\BibitemShut {NoStop}%
\bibitem [{\citenamefont {Ikeda}\ \emph {et~al.}(2010)\citenamefont {Ikeda},
  \citenamefont {Miura}, \citenamefont {Yamamoto}, \citenamefont {Mizunuma},
  \citenamefont {Gan}, \citenamefont {Endo}, \citenamefont {Kanai},
  \citenamefont {Hayakawa}, \citenamefont {Matsukura},\ and\ \citenamefont
  {Ohno}}]{ike10}%
  \BibitemOpen
  \bibfield  {author} {\bibinfo {author} {\bibfnamefont {S.}~\bibnamefont
  {Ikeda}}, \bibinfo {author} {\bibfnamefont {K.}~\bibnamefont {Miura}},
  \bibinfo {author} {\bibfnamefont {H.}~\bibnamefont {Yamamoto}}, \bibinfo
  {author} {\bibfnamefont {K.}~\bibnamefont {Mizunuma}}, \bibinfo {author}
  {\bibfnamefont {H.~D.}\ \bibnamefont {Gan}}, \bibinfo {author} {\bibfnamefont
  {M.}~\bibnamefont {Endo}}, \bibinfo {author} {\bibfnamefont {S.}~\bibnamefont
  {Kanai}}, \bibinfo {author} {\bibfnamefont {J.}~\bibnamefont {Hayakawa}},
  \bibinfo {author} {\bibfnamefont {F.}~\bibnamefont {Matsukura}}, \ and\
  \bibinfo {author} {\bibfnamefont {H.}~\bibnamefont {Ohno}},\ }\href@noop {}
  {\bibfield  {journal} {\bibinfo  {journal} {Nat. Mater.}\ }\textbf {\bibinfo
  {volume} {9}},\ \bibinfo {pages} {721} (\bibinfo {year} {2010})}\BibitemShut
  {NoStop}%
\bibitem [{\citenamefont {Cubukcu}\ \emph {et~al.}(2014)\citenamefont
  {Cubukcu}, \citenamefont {Boulle}, \citenamefont {Drouard}, \citenamefont
  {Garello}, \citenamefont {Avci}, \citenamefont {Miron}, \citenamefont
  {Langer}, \citenamefont {Ocker}, \citenamefont {Gambardella},\ and\
  \citenamefont {Gaudin}}]{cub14}%
  \BibitemOpen
  \bibfield  {author} {\bibinfo {author} {\bibfnamefont {M.}~\bibnamefont
  {Cubukcu}}, \bibinfo {author} {\bibfnamefont {O.}~\bibnamefont {Boulle}},
  \bibinfo {author} {\bibfnamefont {M.}~\bibnamefont {Drouard}}, \bibinfo
  {author} {\bibfnamefont {K.}~\bibnamefont {Garello}}, \bibinfo {author}
  {\bibfnamefont {C.~O.}\ \bibnamefont {Avci}}, \bibinfo {author}
  {\bibfnamefont {I.~M.}\ \bibnamefont {Miron}}, \bibinfo {author}
  {\bibfnamefont {J.}~\bibnamefont {Langer}}, \bibinfo {author} {\bibfnamefont
  {B.}~\bibnamefont {Ocker}}, \bibinfo {author} {\bibfnamefont
  {P.}~\bibnamefont {Gambardella}}, \ and\ \bibinfo {author} {\bibfnamefont
  {G.}~\bibnamefont {Gaudin}},\ }\href@noop {} {\bibfield  {journal} {\bibinfo
  {journal} {Appl. Phys. Lett.}\ }\textbf {\bibinfo {volume} {104}},\ \bibinfo
  {pages} {042406} (\bibinfo {year} {2014})}\BibitemShut {NoStop}%
\bibitem [{\citenamefont {McGarvey}(1966)}]{mcg66}%
  \BibitemOpen
  \bibfield  {author} {\bibinfo {author} {\bibfnamefont {B.}~\bibnamefont
  {McGarvey}},\ }in\ \href@noop {} {\emph {\bibinfo {booktitle} {Transition
  Metal Chemistry}}},\ Vol.~\bibinfo {volume} {3},\ \bibinfo {editor} {edited
  by\ \bibinfo {editor} {\bibfnamefont {R.~L.}\ \bibnamefont {Carlin}}\ and\
  \bibinfo {editor} {\bibfnamefont {M.}~\bibnamefont {Dekker}}}\ (\bibinfo
  {publisher} {John Wiley \& Sons},\ \bibinfo {address} {New York},\ \bibinfo
  {year} {1966})\BibitemShut {NoStop}%
\bibitem [{\citenamefont {Gatteschi}\ \emph {et~al.}(2006)\citenamefont
  {Gatteschi}, \citenamefont {Sessoli},\ and\ \citenamefont {Villain}}]{gat06}%
  \BibitemOpen
  \bibfield  {author} {\bibinfo {author} {\bibfnamefont {D.}~\bibnamefont
  {Gatteschi}}, \bibinfo {author} {\bibfnamefont {R.}~\bibnamefont {Sessoli}},
  \ and\ \bibinfo {author} {\bibfnamefont {J.}~\bibnamefont {Villain}},\
  }\href@noop {} {\emph {\bibinfo {title} {Molecular Nanomagnets}}}\ (\bibinfo
  {publisher} {Oxford University Press},\ \bibinfo {address} {Oxford},\
  \bibinfo {year} {2006})\BibitemShut {NoStop}%
\bibitem [{\citenamefont {Schintke}\ \emph {et~al.}(2001)\citenamefont
  {Schintke}, \citenamefont {Messerli}, \citenamefont {Pivetta}, \citenamefont
  {Patthey}, \citenamefont {Libioulle}, \citenamefont {Stengel}, \citenamefont
  {De~Vita},\ and\ \citenamefont {Schneider}}]{sin01}%
  \BibitemOpen
  \bibfield  {author} {\bibinfo {author} {\bibfnamefont {S.}~\bibnamefont
  {Schintke}}, \bibinfo {author} {\bibfnamefont {S.}~\bibnamefont {Messerli}},
  \bibinfo {author} {\bibfnamefont {M.}~\bibnamefont {Pivetta}}, \bibinfo
  {author} {\bibfnamefont {F.}~\bibnamefont {Patthey}}, \bibinfo {author}
  {\bibfnamefont {L.}~\bibnamefont {Libioulle}}, \bibinfo {author}
  {\bibfnamefont {M.}~\bibnamefont {Stengel}}, \bibinfo {author} {\bibfnamefont
  {A.}~\bibnamefont {De~Vita}}, \ and\ \bibinfo {author} {\bibfnamefont
  {W.~D.}\ \bibnamefont {Schneider}},\ }\href@noop {} {\bibfield  {journal}
  {\bibinfo  {journal} {Phys. Rev. Lett.}\ }\textbf {\bibinfo {volume} {87}},\
  \bibinfo {pages} {276801} (\bibinfo {year} {2001})}\BibitemShut {NoStop}%
\bibitem [{\citenamefont {Neyman}\ \emph {et~al.}(2004)\citenamefont {Neyman},
  \citenamefont {Inntam}, \citenamefont {Nasluzov}, \citenamefont {Kosarev},\
  and\ \citenamefont {R{\"o}sch}}]{ney04}%
  \BibitemOpen
  \bibfield  {author} {\bibinfo {author} {\bibfnamefont {K.}~\bibnamefont
  {Neyman}}, \bibinfo {author} {\bibfnamefont {C.}~\bibnamefont {Inntam}},
  \bibinfo {author} {\bibfnamefont {V.}~\bibnamefont {Nasluzov}}, \bibinfo
  {author} {\bibfnamefont {R.}~\bibnamefont {Kosarev}}, \ and\ \bibinfo
  {author} {\bibfnamefont {N.}~\bibnamefont {R{\"o}sch}},\ }\href@noop {}
  {\bibfield  {journal} {\bibinfo  {journal} {Appl. Phys. A}\ }\textbf
  {\bibinfo {volume} {78}},\ \bibinfo {pages} {823} (\bibinfo {year}
  {2004})}\BibitemShut {NoStop}%
\bibitem [{\citenamefont {Baumann}\ \emph {et~al.}(2014)\citenamefont
  {Baumann}, \citenamefont {Rau}, \citenamefont {Loth}, \citenamefont {Lutz},\
  and\ \citenamefont {Heinrich}}]{bau14}%
  \BibitemOpen
  \bibfield  {author} {\bibinfo {author} {\bibfnamefont {S.}~\bibnamefont
  {Baumann}}, \bibinfo {author} {\bibfnamefont {I.~G.}\ \bibnamefont {Rau}},
  \bibinfo {author} {\bibfnamefont {S.}~\bibnamefont {Loth}}, \bibinfo {author}
  {\bibfnamefont {C.~P.}\ \bibnamefont {Lutz}}, \ and\ \bibinfo {author}
  {\bibfnamefont {A.~J.}\ \bibnamefont {Heinrich}},\ }\href@noop {} {\bibfield
  {journal} {\bibinfo  {journal} {ACS Nano}\ }\textbf {\bibinfo {volume} {8}},\
  \bibinfo {pages} {1739} (\bibinfo {year} {2014})}\BibitemShut {NoStop}%
%
\bibitem [{\citenamefont {Ferr\'{o}n}\ \emph {et~al.}(2015)\citenamefont
  {Ferr\'{o}n}, \citenamefont {Delgado},\ and\ \citenamefont {Fern\'{a}ndez-Rossier}}]{fer15}%
  \BibitemOpen
  \bibfield  {author} {\bibinfo {author} {\bibfnamefont {A.}~\bibnamefont
  {Ferr\'{o}n}}, \bibinfo {author} {\bibfnamefont {F.}\ \bibnamefont {Delgado}},\ and\ \bibinfo {author}
  {\bibfnamefont {J.}\ \bibnamefont {Fern\'{a}ndez-Rossier}},\ }\href@noop {} {\bibfield
  {journal} {\bibinfo  {journal} {N.~J.~Phys.}\ }\textbf {\bibinfo {volume} {17}},\
  \bibinfo {pages} {033020} (\bibinfo {year} {2015})}\BibitemShut {NoStop}%	
%		
\bibitem [{sup()}]{sup}%
  \BibitemOpen
  \href@noop {} {}\bibinfo {note} {See Supplementary Information, which includes Refs. \cite{jac66,gri63,gia09,bla01,sgr01,bro85,eri90,rod01,leh10,van91,gun86}, for details
  concerning sample preparation, XAS and XMCD measurements, and multiplet
  calculations.}\BibitemShut {Stop}%
\bibitem [{\citenamefont {Jaklevic}\ and\ \citenamefont {Lambe}(1966)}]{jac66}%
  \BibitemOpen
  \bibfield  {author} {\bibinfo {author} {\bibfnamefont {R.~C.}\ \bibnamefont
  {Jaklevic}}\ and\ \bibinfo {author} {\bibfnamefont {J.}~\bibnamefont
  {Lambe}},\ }\href@noop {} {\bibfield  {journal} {\bibinfo  {journal} {Phys.
  Rev. Lett.}\ }\textbf {\bibinfo {volume} {17}},\ \bibinfo {pages} {1139}
  (\bibinfo {year} {1966})}\BibitemShut {NoStop}%
\bibitem [{\citenamefont {Griffith}(1963)}]{gri63}%
  \BibitemOpen
  \bibfield  {author} {\bibinfo {author} {\bibfnamefont {J.~S.}\ \bibnamefont
  {Griffith}},\ }\href@noop {} {\bibfield  {journal} {\bibinfo  {journal}
  {Phys. Rev.}\ }\textbf {\bibinfo {volume} {132}},\ \bibinfo {pages} {316}
  (\bibinfo {year} {1963})}\BibitemShut {NoStop}%
\bibitem [{\citenamefont {Giannozzi}\ \emph {et~al.}(2009)\citenamefont
  {Giannozzi}, \citenamefont {Baroni}, \citenamefont {Bonini}, \citenamefont
  {Calandra}, \citenamefont {Car}, \citenamefont {Cavazzoni}, \citenamefont
  {Ceresoli}, \citenamefont {Chiarotti}, \citenamefont {Cococcioni},
  \citenamefont {Dabo}, \citenamefont {Corso}, \citenamefont {de~Gironcoli},
  \citenamefont {Fabris}, \citenamefont {Fratesi}, \citenamefont {Gebauer},
  \citenamefont {Gerstmann}, \citenamefont {Gougoussis}, \citenamefont
  {Kokalj}, \citenamefont {Lazzeri}, \citenamefont {Martin-Samos},
  \citenamefont {Marzari}, \citenamefont {Mauri}, \citenamefont {Mazzarello},
  \citenamefont {Paolini}, \citenamefont {Pasquarello}, \citenamefont
  {Paulatto}, \citenamefont {Sbraccia}, \citenamefont {Scandolo}, \citenamefont
  {Sclauzero}, \citenamefont {Seitsonen}, \citenamefont {Smogunov},
  \citenamefont {Umari},\ and\ \citenamefont {Wentzcovitch}}]{gia09}%
  \BibitemOpen
  \bibfield  {author} {\bibinfo {author} {\bibfnamefont {P.}~\bibnamefont
  {Giannozzi}}, \bibinfo {author} {\bibfnamefont {S.}~\bibnamefont {Baroni}},
  \bibinfo {author} {\bibfnamefont {N.}~\bibnamefont {Bonini}}, \bibinfo
  {author} {\bibfnamefont {M.}~\bibnamefont {Calandra}}, \bibinfo {author}
  {\bibfnamefont {R.}~\bibnamefont {Car}}, \bibinfo {author} {\bibfnamefont
  {C.}~\bibnamefont {Cavazzoni}}, \bibinfo {author} {\bibfnamefont
  {D.}~\bibnamefont {Ceresoli}}, \bibinfo {author} {\bibfnamefont {G.~L.}\
  \bibnamefont {Chiarotti}}, \bibinfo {author} {\bibfnamefont {M.}~\bibnamefont
  {Cococcioni}}, \bibinfo {author} {\bibfnamefont {I.}~\bibnamefont {Dabo}},
  \bibinfo {author} {\bibfnamefont {A.~D.}\ \bibnamefont {Corso}}, \bibinfo
  {author} {\bibfnamefont {S.}~\bibnamefont {de~Gironcoli}}, \bibinfo {author}
  {\bibfnamefont {S.}~\bibnamefont {Fabris}}, \bibinfo {author} {\bibfnamefont
  {G.}~\bibnamefont {Fratesi}}, \bibinfo {author} {\bibfnamefont
  {R.}~\bibnamefont {Gebauer}}, \bibinfo {author} {\bibfnamefont
  {U.}~\bibnamefont {Gerstmann}}, \bibinfo {author} {\bibfnamefont
  {C.}~\bibnamefont {Gougoussis}}, \bibinfo {author} {\bibfnamefont
  {A.}~\bibnamefont {Kokalj}}, \bibinfo {author} {\bibfnamefont
  {M.}~\bibnamefont {Lazzeri}}, \bibinfo {author} {\bibfnamefont
  {L.}~\bibnamefont {Martin-Samos}}, \bibinfo {author} {\bibfnamefont
  {N.}~\bibnamefont {Marzari}}, \bibinfo {author} {\bibfnamefont
  {F.}~\bibnamefont {Mauri}}, \bibinfo {author} {\bibfnamefont
  {R.}~\bibnamefont {Mazzarello}}, \bibinfo {author} {\bibfnamefont
  {S.}~\bibnamefont {Paolini}}, \bibinfo {author} {\bibfnamefont
  {A.}~\bibnamefont {Pasquarello}}, \bibinfo {author} {\bibfnamefont
  {L.}~\bibnamefont {Paulatto}}, \bibinfo {author} {\bibfnamefont
  {C.}~\bibnamefont {Sbraccia}}, \bibinfo {author} {\bibfnamefont
  {S.}~\bibnamefont {Scandolo}}, \bibinfo {author} {\bibfnamefont
  {G.}~\bibnamefont {Sclauzero}}, \bibinfo {author} {\bibfnamefont {A.~P.}\
  \bibnamefont {Seitsonen}}, \bibinfo {author} {\bibfnamefont {A.}~\bibnamefont
  {Smogunov}}, \bibinfo {author} {\bibfnamefont {P.}~\bibnamefont {Umari}}, \
  and\ \bibinfo {author} {\bibfnamefont {R.~M.}\ \bibnamefont {Wentzcovitch}},\
  }\href@noop {} {\bibfield  {journal} {\bibinfo  {journal} {J. Phys. Condens.
  Matter}\ }\textbf {\bibinfo {volume} {21}},\ \bibinfo {pages} {395502}
  (\bibinfo {year} {2009})}\BibitemShut {NoStop}%
\bibitem [{\citenamefont {Blaha}\ \emph {et~al.}(2001)\citenamefont {Blaha},
  \citenamefont {Schwarz}, \citenamefont {Madsen}, \citenamefont {Kvasnicka},\
  and\ \citenamefont {Luitz}}]{bla01}%
  \BibitemOpen
  \bibfield  {author} {\bibinfo {author} {\bibfnamefont {P.}~\bibnamefont
  {Blaha}}, \bibinfo {author} {\bibfnamefont {K.}~\bibnamefont {Schwarz}},
  \bibinfo {author} {\bibfnamefont {G.}~\bibnamefont {Madsen}}, \bibinfo
  {author} {\bibfnamefont {D.}~\bibnamefont {Kvasnicka}}, \ and\ \bibinfo
  {author} {\bibfnamefont {J.}~\bibnamefont {Luitz}},\ }\href@noop {} {\emph
  {\bibinfo {title} {WIEN2k: An augmented plane wave+ local orbitals program
  for calculating crystal properties.}}}\ (\bibinfo  {publisher} {Karlheinz
  Schwarz, Techn. Universitat Wien, Austria},\ \bibinfo {year}
  {2001})\BibitemShut {NoStop}%
\bibitem [{\citenamefont {Sgroi}\ \emph {et~al.}(2001)\citenamefont {Sgroi},
  \citenamefont {Pisani},\ and\ \citenamefont {Busso}}]{sgr01}%
  \BibitemOpen
  \bibfield  {author} {\bibinfo {author} {\bibfnamefont {M.}~\bibnamefont
  {Sgroi}}, \bibinfo {author} {\bibfnamefont {C.}~\bibnamefont {Pisani}}, \
  and\ \bibinfo {author} {\bibfnamefont {M.}~\bibnamefont {Busso}},\
  }\href@noop {} {\bibfield  {journal} {\bibinfo  {journal} {Thin Solid Films}\
  }\textbf {\bibinfo {volume} {400}},\ \bibinfo {pages} {64} (\bibinfo {year}
  {2001})}\BibitemShut {NoStop}%
\bibitem [{\citenamefont {Brooks}(1985)}]{bro85}%
  \BibitemOpen
  \bibfield  {author} {\bibinfo {author} {\bibfnamefont {M.}~\bibnamefont
  {Brooks}},\ }\href@noop {} {\bibfield  {journal} {\bibinfo  {journal}
  {Physica B+C}\ }\textbf {\bibinfo {volume} {130}},\ \bibinfo {pages} {6 }
  (\bibinfo {year} {1985})}\BibitemShut {NoStop}%
\bibitem [{\citenamefont {Eriksson}\ \emph {et~al.}(1990)\citenamefont
  {Eriksson}, \citenamefont {Johansson}, \citenamefont {Albers}, \citenamefont
  {Boring},\ and\ \citenamefont {Brooks}}]{eri90}%
  \BibitemOpen
  \bibfield  {author} {\bibinfo {author} {\bibfnamefont {O.}~\bibnamefont
  {Eriksson}}, \bibinfo {author} {\bibfnamefont {B.}~\bibnamefont {Johansson}},
  \bibinfo {author} {\bibfnamefont {R.~C.}\ \bibnamefont {Albers}}, \bibinfo
  {author} {\bibfnamefont {A.~M.}\ \bibnamefont {Boring}}, \ and\ \bibinfo
  {author} {\bibfnamefont {M.~S.~S.}\ \bibnamefont {Brooks}},\ }\href@noop {}
  {\bibfield  {journal} {\bibinfo  {journal} {Phys. Rev. B}\ }\textbf {\bibinfo
  {volume} {42}},\ \bibinfo {pages} {2707} (\bibinfo {year}
  {1990})}\BibitemShut {NoStop}%
\bibitem [{\citenamefont {Rodriguez}\ \emph {et~al.}(2001)\citenamefont
  {Rodriguez}, \citenamefont {Ganduglia-Pirovano}, \citenamefont {Peltzer~y
  Blanc\'a}, \citenamefont {Petersen},\ and\ \citenamefont {Nov\'ak}}]{rod01}%
  \BibitemOpen
  \bibfield  {author} {\bibinfo {author} {\bibfnamefont {C.~O.}\ \bibnamefont
  {Rodriguez}}, \bibinfo {author} {\bibfnamefont {M.~V.}\ \bibnamefont
  {Ganduglia-Pirovano}}, \bibinfo {author} {\bibfnamefont {E.~L.}\ \bibnamefont
  {Peltzer~y Blanc\'a}}, \bibinfo {author} {\bibfnamefont {M.}~\bibnamefont
  {Petersen}}, \ and\ \bibinfo {author} {\bibfnamefont {P.}~\bibnamefont
  {Nov\'ak}},\ }\href@noop {} {\bibfield  {journal} {\bibinfo  {journal} {Phys.
  Rev. B}\ }\textbf {\bibinfo {volume} {63}},\ \bibinfo {pages} {184413}
  (\bibinfo {year} {2001})}\BibitemShut {NoStop}%
\bibitem [{\citenamefont {Lehnert}\ \emph {et~al.}(2010)\citenamefont
  {Lehnert}, \citenamefont {Rusponi}, \citenamefont {Etzkorn}, \citenamefont
  {Ouazi}, \citenamefont {Thakur},\ and\ \citenamefont {Brune}}]{leh10}%
  \BibitemOpen
  \bibfield  {author} {\bibinfo {author} {\bibfnamefont {A.}~\bibnamefont
  {Lehnert}}, \bibinfo {author} {\bibfnamefont {S.}~\bibnamefont {Rusponi}},
  \bibinfo {author} {\bibfnamefont {M.}~\bibnamefont {Etzkorn}}, \bibinfo
  {author} {\bibfnamefont {S.}~\bibnamefont {Ouazi}}, \bibinfo {author}
  {\bibfnamefont {P.}~\bibnamefont {Thakur}}, \ and\ \bibinfo {author}
  {\bibfnamefont {H.}~\bibnamefont {Brune}},\ }\href@noop {} {\bibfield
  {journal} {\bibinfo  {journal} {Phys. Rev. B}\ }\textbf {\bibinfo {volume}
  {81}},\ \bibinfo {pages} {104430} (\bibinfo {year} {2010})}\BibitemShut
  {NoStop}%
\bibitem [{\citenamefont {van~der Laan}\ and\ \citenamefont
  {Thole}(1991)}]{van91}%
  \BibitemOpen
  \bibfield  {author} {\bibinfo {author} {\bibfnamefont {G.}~\bibnamefont
  {van~der Laan}}\ and\ \bibinfo {author} {\bibfnamefont {B.~T.}\ \bibnamefont
  {Thole}},\ }\href@noop {} {\bibfield  {journal} {\bibinfo  {journal} {Phys.
  Rev. B}\ }\textbf {\bibinfo {volume} {43}},\ \bibinfo {pages} {13401}
  (\bibinfo {year} {1991})}\BibitemShut {NoStop}%
\bibitem [{\citenamefont {Gunnarsson}\ and\ \citenamefont
  {Sch\"onhammer}(1983)}]{gun86}%
  \BibitemOpen
  \bibfield  {author} {\bibinfo {author} {\bibfnamefont {O.}~\bibnamefont
  {Gunnarsson}}\ and\ \bibinfo {author} {\bibfnamefont {K.}~\bibnamefont
  {Sch\"onhammer}},\ }\href@noop {} {\bibfield  {journal} {\bibinfo  {journal}
  {Phys. Rev. B}\ }\textbf {\bibinfo {volume} {28}},\ \bibinfo {pages} {4315}
  (\bibinfo {year} {1983})}\BibitemShut {NoStop}%
\bibitem [{\citenamefont {Cococcioni}\ and\ \citenamefont
  {de~Gironcoli}(2005)}]{coc05}%
  \BibitemOpen
  \bibfield  {author} {\bibinfo {author} {\bibfnamefont {M.}~\bibnamefont
  {Cococcioni}}\ and\ \bibinfo {author} {\bibfnamefont {S.}~\bibnamefont
  {de~Gironcoli}},\ }\href@noop {} {\bibfield  {journal} {\bibinfo  {journal}
  {Phys. Rev. B}\ }\textbf {\bibinfo {volume} {71}},\ \bibinfo {pages} {035105}
  (\bibinfo {year} {2005})}\BibitemShut {NoStop}%
\bibitem [{\citenamefont {Piamonteze}\ \emph {et~al.}(2012)\citenamefont
  {Piamonteze}, \citenamefont {Flechsig}, \citenamefont {Rusponi},
  \citenamefont {Dreiser}, \citenamefont {Heidler}, \citenamefont {Schmidt},
  \citenamefont {Wetter}, \citenamefont {Calvi}, \citenamefont {Schmidt},
  \citenamefont {Pruchova}, \citenamefont {Krempasky}, \citenamefont
  {Quitmann}, \citenamefont {Brune},\ and\ \citenamefont {Nolting}}]{pia12}%
  \BibitemOpen
  \bibfield  {author} {\bibinfo {author} {\bibfnamefont {C.}~\bibnamefont
  {Piamonteze}}, \bibinfo {author} {\bibfnamefont {U.}~\bibnamefont
  {Flechsig}}, \bibinfo {author} {\bibfnamefont {S.}~\bibnamefont {Rusponi}},
  \bibinfo {author} {\bibfnamefont {J.}~\bibnamefont {Dreiser}}, \bibinfo
  {author} {\bibfnamefont {J.}~\bibnamefont {Heidler}}, \bibinfo {author}
  {\bibfnamefont {M.}~\bibnamefont {Schmidt}}, \bibinfo {author} {\bibfnamefont
  {R.}~\bibnamefont {Wetter}}, \bibinfo {author} {\bibfnamefont
  {M.}~\bibnamefont {Calvi}}, \bibinfo {author} {\bibfnamefont
  {T.}~\bibnamefont {Schmidt}}, \bibinfo {author} {\bibfnamefont
  {H.}~\bibnamefont {Pruchova}}, \bibinfo {author} {\bibfnamefont
  {J.}~\bibnamefont {Krempasky}}, \bibinfo {author} {\bibfnamefont
  {C.}~\bibnamefont {Quitmann}}, \bibinfo {author} {\bibfnamefont
  {H.}~\bibnamefont {Brune}}, \ and\ \bibinfo {author} {\bibfnamefont
  {F.}~\bibnamefont {Nolting}},\ }\href@noop {} {\bibfield  {journal} {\bibinfo
   {journal} {J. Synchrotron Rad.}\ }\textbf {\bibinfo {volume} {19}},\
  \bibinfo {pages} {661} (\bibinfo {year} {2012})}\BibitemShut {NoStop}%
\bibitem [{\citenamefont {Gambardella}\ \emph {et~al.}(2002)\citenamefont
  {Gambardella}, \citenamefont {Dhesi}, \citenamefont {Gardonio}, \citenamefont
  {Grazioli}, \citenamefont {Ohresser},\ and\ \citenamefont {Carbone}}]{gam02}%
  \BibitemOpen
  \bibfield  {author} {\bibinfo {author} {\bibfnamefont {P.}~\bibnamefont
  {Gambardella}}, \bibinfo {author} {\bibfnamefont {S.~S.}\ \bibnamefont
  {Dhesi}}, \bibinfo {author} {\bibfnamefont {S.}~\bibnamefont {Gardonio}},
  \bibinfo {author} {\bibfnamefont {C.}~\bibnamefont {Grazioli}}, \bibinfo
  {author} {\bibfnamefont {P.}~\bibnamefont {Ohresser}}, \ and\ \bibinfo
  {author} {\bibfnamefont {C.}~\bibnamefont {Carbone}},\ }\href@noop {}
  {\bibfield  {journal} {\bibinfo  {journal} {Phys. Rev. Lett.}\ }\textbf
  {\bibinfo {volume} {88}},\ \bibinfo {pages} {047202} (\bibinfo {year}
  {2002})}\BibitemShut {NoStop}%
\bibitem [{\citenamefont {Thole}\ \emph {et~al.}(1992)\citenamefont {Thole},
  \citenamefont {Carra}, \citenamefont {Sette},\ and\ \citenamefont {van~der
  Laan}}]{tho92}%
  \BibitemOpen
  \bibfield  {author} {\bibinfo {author} {\bibfnamefont {B.~T.}\ \bibnamefont
  {Thole}}, \bibinfo {author} {\bibfnamefont {P.}~\bibnamefont {Carra}},
  \bibinfo {author} {\bibfnamefont {F.}~\bibnamefont {Sette}}, \ and\ \bibinfo
  {author} {\bibfnamefont {G.}~\bibnamefont {van~der Laan}},\ }\href@noop {}
  {\bibfield  {journal} {\bibinfo  {journal} {Phys. Rev. Lett.}\ }\textbf
  {\bibinfo {volume} {68}},\ \bibinfo {pages} {1943} (\bibinfo {year}
  {1992})}\BibitemShut {NoStop}%
\bibitem [{\citenamefont {Carra}\ \emph {et~al.}(1993)\citenamefont {Carra},
  \citenamefont {Thole}, \citenamefont {Altarelli},\ and\ \citenamefont
  {Wang}}]{car93}%
  \BibitemOpen
  \bibfield  {author} {\bibinfo {author} {\bibfnamefont {P.}~\bibnamefont
  {Carra}}, \bibinfo {author} {\bibfnamefont {B.~T.}\ \bibnamefont {Thole}},
  \bibinfo {author} {\bibfnamefont {M.}~\bibnamefont {Altarelli}}, \ and\
  \bibinfo {author} {\bibfnamefont {X.}~\bibnamefont {Wang}},\ }\href@noop {}
  {\bibfield  {journal} {\bibinfo  {journal} {Phys. Rev. Lett.}\ }\textbf
  {\bibinfo {volume} {70}},\ \bibinfo {pages} {694} (\bibinfo {year}
  {1993})}\BibitemShut {NoStop}%
\bibitem [{\citenamefont {Chen}\ \emph {et~al.}(1995)\citenamefont {Chen},
  \citenamefont {Idzerda}, \citenamefont {Lin}, \citenamefont {Smith},
  \citenamefont {Meigs}, \citenamefont {Chaban}, \citenamefont {Ho},
  \citenamefont {Pellegrin},\ and\ \citenamefont {Sette}}]{che95}%
  \BibitemOpen
  \bibfield  {author} {\bibinfo {author} {\bibfnamefont {C.~T.}\ \bibnamefont
  {Chen}}, \bibinfo {author} {\bibfnamefont {Y.~U.}\ \bibnamefont {Idzerda}},
  \bibinfo {author} {\bibfnamefont {H.-J.}\ \bibnamefont {Lin}}, \bibinfo
  {author} {\bibfnamefont {N.~V.}\ \bibnamefont {Smith}}, \bibinfo {author}
  {\bibfnamefont {G.}~\bibnamefont {Meigs}}, \bibinfo {author} {\bibfnamefont
  {E.}~\bibnamefont {Chaban}}, \bibinfo {author} {\bibfnamefont {G.~H.}\
  \bibnamefont {Ho}}, \bibinfo {author} {\bibfnamefont {E.}~\bibnamefont
  {Pellegrin}}, \ and\ \bibinfo {author} {\bibfnamefont {F.}~\bibnamefont
  {Sette}},\ }\href@noop {} {\bibfield  {journal} {\bibinfo  {journal} {Phys.
  Rev. Lett.}\ }\textbf {\bibinfo {volume} {75}},\ \bibinfo {pages} {152}
  (\bibinfo {year} {1995})}\BibitemShut {NoStop}%
\bibitem [{\citenamefont {de~Groot}(2001)}]{gro01}%
  \BibitemOpen
  \bibfield  {author} {\bibinfo {author} {\bibfnamefont {F.}~\bibnamefont
  {de~Groot}},\ }\href@noop {} {\bibfield  {journal} {\bibinfo  {journal}
  {Chem. Rev.}\ }\textbf {\bibinfo {volume} {101}},\ \bibinfo {pages} {1779}
  (\bibinfo {year} {2001})}\BibitemShut {NoStop}%
\bibitem [{mae()}]{mae}%
  \BibitemOpen
  \href@noop {} {}\bibinfo {note} {The magnetic anisotropy energy is the
  activation energy for magnetization reversal, given in our case by $E_{| 4
  \rangle} - E_{| 0 \rangle} = 19$~meV.}\BibitemShut {Stop}%
\bibitem [{\citenamefont {Zadrozny}\ \emph {et~al.}(2013)\citenamefont
  {Zadrozny}, \citenamefont {Xia}, \citenamefont {Atanasov}, \citenamefont
  {Long}, \citenamefont {Grandjean}, \citenamefont {Neese},\ and\ \citenamefont
  {Long}}]{zad13}%
  \BibitemOpen
  \bibfield  {author} {\bibinfo {author} {\bibfnamefont {J.~M.}\ \bibnamefont
  {Zadrozny}}, \bibinfo {author} {\bibfnamefont {D.~J.}\ \bibnamefont {Xia}},
  \bibinfo {author} {\bibfnamefont {M.}~\bibnamefont {Atanasov}}, \bibinfo
  {author} {\bibfnamefont {G.~J.}\ \bibnamefont {Long}}, \bibinfo {author}
  {\bibfnamefont {F.}~\bibnamefont {Grandjean}}, \bibinfo {author}
  {\bibfnamefont {F.}~\bibnamefont {Neese}}, \ and\ \bibinfo {author}
  {\bibfnamefont {J.~R.}\ \bibnamefont {Long}},\ }\href@noop {} {\bibfield
  {journal} {\bibinfo  {journal} {Nat. Chem.}\ }\textbf {\bibinfo {volume}
  {5}},\ \bibinfo {pages} {577} (\bibinfo {year} {2013})}\BibitemShut {NoStop}%
\bibitem [{\citenamefont {Loth}\ \emph
  {et~al.}(2010{\natexlab{a}})\citenamefont {Loth}, \citenamefont {von
  Bergmann}, \citenamefont {Ternes}, \citenamefont {Otte}, \citenamefont
  {Lutz},\ and\ \citenamefont {Heinrich}}]{lot10a}%
  \BibitemOpen
  \bibfield  {author} {\bibinfo {author} {\bibfnamefont {S.}~\bibnamefont
  {Loth}}, \bibinfo {author} {\bibfnamefont {K.}~\bibnamefont {von Bergmann}},
  \bibinfo {author} {\bibfnamefont {M.}~\bibnamefont {Ternes}}, \bibinfo
  {author} {\bibfnamefont {A.~F.}\ \bibnamefont {Otte}}, \bibinfo {author}
  {\bibfnamefont {C.~P.}\ \bibnamefont {Lutz}}, \ and\ \bibinfo {author}
  {\bibfnamefont {A.~J.}\ \bibnamefont {Heinrich}},\ }\href@noop {} {\bibfield
  {journal} {\bibinfo  {journal} {Nat. Phys.}\ }\textbf {\bibinfo {volume}
  {6}},\ \bibinfo {pages} {340} (\bibinfo {year}
  {2010}{\natexlab{a}})}\BibitemShut {NoStop}%
\bibitem [{\citenamefont {Chilian}(2011)}]{chi11}%
  \BibitemOpen
  \bibfield  {author} {\bibinfo {author} {\bibfnamefont {B.}~\bibnamefont
  {Chilian}},\ }\href@noop {} {\bibfield  {journal} {\bibinfo  {journal} {Phys.
  Rev. B}\ }\textbf {\bibinfo {volume} {84}},\ \bibinfo {pages} {212401}
  (\bibinfo {year} {2011})}\BibitemShut {NoStop}%
\bibitem [{\citenamefont {Schuh}\ \emph {et~al.}(2011)\citenamefont {Schuh},
  \citenamefont {Balashov}, \citenamefont {Miyamachi}, \citenamefont {Wu},
  \citenamefont {Kuo}, \citenamefont {Ernst}, \citenamefont {Henk},\ and\
  \citenamefont {Wulfhekel}}]{suh11}%
  \BibitemOpen
  \bibfield  {author} {\bibinfo {author} {\bibfnamefont {T.}~\bibnamefont
  {Schuh}}, \bibinfo {author} {\bibfnamefont {T.}~\bibnamefont {Balashov}},
  \bibinfo {author} {\bibfnamefont {T.}~\bibnamefont {Miyamachi}}, \bibinfo
  {author} {\bibfnamefont {S.~Y.}\ \bibnamefont {Wu}}, \bibinfo {author}
  {\bibfnamefont {C.~C.}\ \bibnamefont {Kuo}}, \bibinfo {author} {\bibfnamefont
  {A.}~\bibnamefont {Ernst}}, \bibinfo {author} {\bibfnamefont
  {J.}~\bibnamefont {Henk}}, \ and\ \bibinfo {author} {\bibfnamefont
  {W.}~\bibnamefont {Wulfhekel}},\ }\href@noop {} {\bibfield  {journal}
  {\bibinfo  {journal} {Phys. Rev. B}\ }\textbf {\bibinfo {volume} {84}},\
  \bibinfo {pages} {104401} (\bibinfo {year} {2011})}\BibitemShut {NoStop}%
\bibitem [{\citenamefont {Abragam}\ and\ \citenamefont
  {Bleaney}(1986)}]{abr86}%
  \BibitemOpen
  \bibfield  {author} {\bibinfo {author} {\bibfnamefont {A.}~\bibnamefont
  {Abragam}}\ and\ \bibinfo {author} {\bibfnamefont {B.}~\bibnamefont
  {Bleaney}},\ }\href@noop {} {\emph {\bibinfo {title} {Electron Paramagnetic
  Resonance of Transition Ions}}}\ (\bibinfo  {publisher} {Clarendon Press},\
  \bibinfo {address} {Oxford},\ \bibinfo {year} {1986})\BibitemShut {NoStop}%
\bibitem [{\citenamefont {Lorente}\ and\ \citenamefont
  {Gauyacq}(2009)}]{lor09}%
  \BibitemOpen
  \bibfield  {author} {\bibinfo {author} {\bibfnamefont {N.}~\bibnamefont
  {Lorente}}\ and\ \bibinfo {author} {\bibfnamefont {J.~P.}\ \bibnamefont
  {Gauyacq}},\ }\href@noop {} {\bibfield  {journal} {\bibinfo  {journal} {Phys.
  Rev. Lett.}\ }\textbf {\bibinfo {volume} {103}},\ \bibinfo {pages} {176601}
  (\bibinfo {year} {2009})}\BibitemShut {NoStop}%
\bibitem [{\citenamefont {Loth}\ \emph
  {et~al.}(2010{\natexlab{b}})\citenamefont {Loth}, \citenamefont {Lutz},\ and\
  \citenamefont {Heinrich}}]{lot10b}%
  \BibitemOpen
  \bibfield  {author} {\bibinfo {author} {\bibfnamefont {S.}~\bibnamefont
  {Loth}}, \bibinfo {author} {\bibfnamefont {C.~P.}\ \bibnamefont {Lutz}}, \
  and\ \bibinfo {author} {\bibfnamefont {A.~J.}\ \bibnamefont {Heinrich}},\
  }\href@noop {} {\bibfield  {journal} {\bibinfo  {journal} {N. J. Phys.}\
  }\textbf {\bibinfo {volume} {12}},\ \bibinfo {pages} {125021} (\bibinfo
  {year} {2010}{\natexlab{b}})}\BibitemShut {NoStop}%
\bibitem [{v06()}]{v06}%
  \BibitemOpen
  \href@noop {} {}\bibinfo {note} {Note that $\Delta S_z = \pm 1, \, 0$ also
  allows the $| 0 \rangle \rightarrow | 6 \rangle$ transition. This transition
  has a higher energy. The fact that it is not giving a significant
  contribution to the $dI/dV$ spectrum signifies that its rate is low and that
  the excited states are long-lived such that we are not limited to transitions
  starting from the ground state.}\BibitemShut {Stop}%
\bibitem [{\citenamefont {Hirjibehedin}\ \emph {et~al.}(2006)\citenamefont
  {Hirjibehedin}, \citenamefont {Lutz},\ and\ \citenamefont
  {Heinrich}}]{hir06}%
  \BibitemOpen
  \bibfield  {author} {\bibinfo {author} {\bibfnamefont {C.~F.}\ \bibnamefont
  {Hirjibehedin}}, \bibinfo {author} {\bibfnamefont {C.~P.}\ \bibnamefont
  {Lutz}}, \ and\ \bibinfo {author} {\bibfnamefont {A.~J.}\ \bibnamefont
  {Heinrich}},\ }\href@noop {} {\bibfield  {journal} {\bibinfo  {journal}
  {Science}\ }\textbf {\bibinfo {volume} {312}},\ \bibinfo {pages} {1021}
  (\bibinfo {year} {2006})}\BibitemShut {NoStop}%
\bibitem [{\citenamefont {Kahle}\ \emph {et~al.}(2012)\citenamefont {Kahle},
  \citenamefont {Deng}, \citenamefont {Malinowski}, \citenamefont {Tonnoir},
  \citenamefont {Forment-Aliaga}, \citenamefont {Thontasen}, \citenamefont
  {Rinke}, \citenamefont {Le}, \citenamefont {Turkowski}, \citenamefont
  {Rahman}, \citenamefont {Rauschenbach}, \citenamefont {Ternes},\ and\
  \citenamefont {Kern}}]{kah12}%
  \BibitemOpen
  \bibfield  {author} {\bibinfo {author} {\bibfnamefont {S.}~\bibnamefont
  {Kahle}}, \bibinfo {author} {\bibfnamefont {Z.}~\bibnamefont {Deng}},
  \bibinfo {author} {\bibfnamefont {N.}~\bibnamefont {Malinowski}}, \bibinfo
  {author} {\bibfnamefont {C.}~\bibnamefont {Tonnoir}}, \bibinfo {author}
  {\bibfnamefont {A.}~\bibnamefont {Forment-Aliaga}}, \bibinfo {author}
  {\bibfnamefont {N.}~\bibnamefont {Thontasen}}, \bibinfo {author}
  {\bibfnamefont {G.}~\bibnamefont {Rinke}}, \bibinfo {author} {\bibfnamefont
  {D.}~\bibnamefont {Le}}, \bibinfo {author} {\bibfnamefont {V.}~\bibnamefont
  {Turkowski}}, \bibinfo {author} {\bibfnamefont {T.~S.}\ \bibnamefont
  {Rahman}}, \bibinfo {author} {\bibfnamefont {S.}~\bibnamefont
  {Rauschenbach}}, \bibinfo {author} {\bibfnamefont {M.}~\bibnamefont
  {Ternes}}, \ and\ \bibinfo {author} {\bibfnamefont {K.}~\bibnamefont
  {Kern}},\ }\href@noop {} {\bibfield  {journal} {\bibinfo  {journal} {Nano
  Lett.}\ }\textbf {\bibinfo {volume} {12}},\ \bibinfo {pages} {518} (\bibinfo
  {year} {2012})}\BibitemShut {NoStop}%
\end{thebibliography}

%

\end{document}